\documentclass[aps,prl,superscriptaddress,preprintnumbers,
showpacs,legalpaper,twoside,twocolumn]{revtex4}

\usepackage{bm}
\usepackage{graphicx}
\usepackage{amsfonts}
\usepackage{amsmath}
\usepackage{amssymb}

\begin{document}

\title{The disordered-free-moment phase: a low-field disordered state in spin-gap
antiferromagnets with site dilution}

\author{Rong Yu}
\affiliation{Department of Physics and Astronomy, University of Southern
California, Los Angeles, CA 90089-0484}
\author{Tommaso Roscilde}
\affiliation{Max-Planck-Institut f\"ur Quantenoptik, Hans-Kopfermann-strasse 1,
85748 Garching, Germany}
\author{Stephan Haas}
\affiliation{Department of Physics and Astronomy, University of Southern
California, Los Angeles, CA 90089-0484}


\begin{abstract}
Site dilution of spin-gapped antiferromagnets leads to
localized free moments, which can order antiferromagnetically in two and
higher dimensions. Here we show how a weak magnetic field drives this
order-by-disorder state into a novel \emph{disordered-free-moment} phase,
characterized by the formation of local singlets between neighboring
moments and by localized moments aligned antiparallel to the field. This
disordered phase is characterized by
the absence of a gap, as it is the case in a Bose glass.
The associated field-driven quantum phase transition is consistent with the
universality of a superfluid-to-Bose-glass transition. The
robustness of the disordered-free-moment phase and its prominent features, in
particular a series of \emph{pseudo}-plateaus in the magnetization
curve, makes it accessible and relevant to experiments.
\end{abstract}
\maketitle

 Valence bond solids (VBS) in spin-gapped
antiferromagnets represent some of the most fundamental
examples of quantum-disordered states
in condensed matter systems. The nature
of such states is by now
well understood theoretically and has been extensively
verified experimentally.
A variety of mechanisms, such as
increased strength of certain bonds in
the magnetic Hamiltonian \cite{SandvikS94, Matsumotoetal01},
magnetic frustration
\cite{ShastryS81}, and the Haldane mechanism \cite{Affleck89},
can render classical
N\'eel order unstable towards the formation of local
singlets, which arrange themselves into a VBS.
Evidence for such phases has been found in a large number
of magnetic compounds, ranging from Haldane chains
\cite{Regnaultetal94}, to spin ladders
\cite{DagottoR96}, to weakly coupled dimer systems
\cite{Cavadinietal00,Sasagoetal97}.
 A central focus of theoretical and experimental
 investigations has been the effect of doping on
 such states.
In particular, it was soon realized theoretically,
 \cite{ShenderK91, SigristF96} and observed
 experimentally \cite{Azumaetal97,Xuetal00}, that doping
 a VBS with static,
 non-magnetic impurities leads to the intriguing
 phenomenon of \emph{order-by-disorder} (OBD):
 free $S=1/2$ magnetic moments
 appear close to the impurity sites and
 interact effectively via a long-range network
 of unfrustrated (albeit random) couplings, which
 decay exponentially with the inter-moment
 distance. These interactions, although weak, are sufficient
 for the free moments (FMs) to order antiferromagnetically
 at experimentally relevant temperatures
 \cite{Azumaetal97}.

\begin{figure}[h]
\begin{center}
\includegraphics[
bbllx=0pt,bblly=0pt,bburx=591pt,bbury=273pt,%
     width=90mm,angle=0]{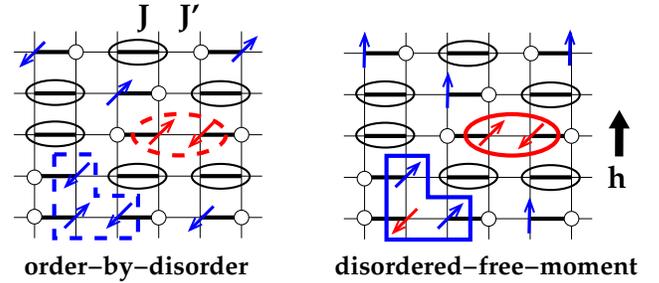}
     \vskip -0.5cm
\caption{(Color online) Quantum phase transition between the
order-by-disorder (OBD) phase and the novel disordered-free-moment
(DFM) phase in a diluted coupled-dimer system. \emph{Left panel}: At
zero applied field, spins on intact dimers form
singlets (solid ellipses), dimers of free moments (FMs) have a strong
singlet component (dashed ellipse), whereas the other FMs (blue
arrows) participate in the OBD state. \emph{Right panel}: Upon
applying a field, the OBD-FMs are mostly polarized, but local
singlets and localized down-spins can survive on clustered FMs, leading
to the DFM phase.} \label{f.DFM}
\end{center}
\null\vspace*{-1.5cm}
\end{figure}

  Given the random nature of the inter-moment
 couplings, the OBD state induced by
 doping is extremely inhomogeneous, as it
 contains a large variety of energy scales which
 depend exponentially on the spatial distribution of the impurities.
 In this paper we study the evolution of
 the OBD state
 upon application of a magnetic field,
 which represents a straightforward experimental
 probe of energy scales in magnetic systems.
 Precisely due to the large distribution
 of the effective couplings between the FMs,
 we find an amazingly rich
 response of the system to the applied
 field. The field scan reveals that the
 long-range order in the system is
 extremely tenuous for large spin gaps,
 and its field-driven destruction leaves behind
 local singlets (or spins oppositely polarized
 to the field) on even- (or odd-)numbered
 clusters of FMs, which are coupled at
 energies higher than the scale characteristic
 for N\'eel order (see Fig. \ref{f.DFM}). The local
 polarization fields for these FM clusters
 cover a continuous range, so that the magnetization
 process with increasing field continues also
 after destruction of the OBD, and the resulting
 \emph{disordered-free-moment} (DFM) phase
 \cite{Roscilde06} is \emph{gapless}. The DFM phase
 persists up to the sizable field which polarizes
 all the FMs, and the magnetization curve within this phase
 shows prominent features of intermediate
 \emph{pseudo}-plateaus, still retaining
 a finite albeit extremely small slope, related
 to the distribution of the strongly interacting
 clusters of FMs.

  To quantitatively investigate the field response
 of a site-diluted spin-gapped antiferromagnet, we
 focus our attention on a two-dimensional $S=1/2$
 model of weakly coupled dimers \cite{Matsumotoetal01,
 Yasudaetal01},
 whose Hamiltonian reads

\begin{eqnarray}
H&=& J\sum_{i\in A} \epsilon_{i}\epsilon_{i+\hat x}
\bm{S}_{i}\cdot
\bm{S}_{i+\hat x}
+J'\sum_{i\in A}\epsilon_{i}\epsilon_{i+\hat y}
\bm{S}_{i}\cdot
\bm{S}_{i+\hat y} \nonumber\\
&+&J'\sum_{i\in B}\sum_{\hat{d} = \hat{x},\hat{y}}
\epsilon_{i}\epsilon_{i+\hat d}
\bm{S}_{i}\cdot \bm{S}_{i+\hat d}
-h \sum_{i}\epsilon_{i} \emph{S}^z_{i}.
\label{e.ham_dilute}
\end{eqnarray}

 Here $i$ runs over the two sublattices ($A$ and $B$)
 of a square lattice, $\hat x$ and $\hat y$ are the
 two lattice vectors, and $\epsilon_i$ is the
 random dilution variable taking values 0 and 1
 with probability $p$ and $1-p$ respectively.
 $h = g\mu_B H$ is the applied field.
 The couplings $J>J'$ determine the subset of
 strong antiferromagnetic bonds: for $J'/J < 0.523..$
 \cite{Matsumotoetal01} the bond anisotropy stabilizes
 a dimer-singlet ground state against the conventional
 N\'eel ordered state of the square-lattice antiferromagnet.
 All the results presented here refer to the field
 and doping effects deep within the dimer-singlet
 regime at $J'/J = 1/4$.

 The presence of lattice vacancies induces
 FMs which are localized in the vicinity of
 the unpaired spins which have lost their
 $J$-neighbor. Perturbation theory
 \cite{SigristF96, Mikeskaetal04, Roscilde06} provides
 an effective coupling $J_{ij} \approx (-1)^{i-j-1} (J_1/r) \exp(-r/\xi_0)$
 between these FMs,
 where $r=|i-j|$, $\xi_0$ is the correlation
 length of the undoped system. We choose $J_1 \approx J' \exp(1/\xi_0)$
 in order for $J_{ij}$ to correctly reproduce the limit
 $J'$ for neighboring unpaired spins. For a deeper understanding of
 the Hamiltonian Eq. (\ref{e.ham_dilute}), it is
 illuminating \cite{Laflorencieetal04} to study an effective
 model for the network of FMs, consisting
 of randomly distributed $S=1/2$ spins with effective couplings $J_{ij}$
 \begin{equation}
 H_{\rm FM} = \frac{1}{2} \sum_{i,j} J_{ij} \bm{S}_{i}\cdot
\bm{S}_{j} - h \sum_{i} \emph{S}^z_{i}.
 \label{e.ham_effective}
 \end{equation}

 We investigate the original and the effective
 Hamiltonian, given by Eq. (\ref{e.ham_dilute}) and
 (\ref{e.ham_effective}), using Stochastic Series Expansion
 (SSE) Quantum Monte Carlo simulations based on the directed-loop
 algorithm \cite{SyljuasenS02}. For the original Hamiltonian
 Eq. (\ref{e.ham_dilute}) we study $L\times L$ lattices
 up to $L$=40 with dilution $p=1/8$, whereas for the effective
 model Eq. (\ref{e.ham_effective}) we randomly distribute
 spins on the same lattice sizes with a density $p$ equal to
 that of the vacancies in the original model.
 Disorder averaging is typically performed over
 $\approx 300$ realizations.
  The ground-state properties are systematically
  obtained using a $\beta$-doubling approach \cite{Sandvik02}.
  Inverse temperatures up to $\beta J = 2^{15}$ are necessary to
  observe the physical $T\rightarrow 0$ behavior. In the following,
  we focus our attention on the uniform
  magnetization per spin
  $m_u = 1/N_s \sum_i \langle S_i^z \rangle$, 
  where $N_s$ is the total number of spins in the system considered;
  on the uniform susceptibility
  $\chi_u/J = \partial m_u / \partial h$;
  on the staggered magnetization
  $m_s = \sqrt{S^{\perp}(\pi,\pi)/L^2}$,
  where $S^{\perp}(\pi,\pi) = 1/(2L^2) \sum_{ij}
  \langle S_i^x S_j^x + S_i^y S_j^y \rangle$ is
  the transverse static structure factor; on the
  correlation length $\xi$, extracted from the
  $q$-dependent structure factor;
  and on the superfluid density $\rho_s= 1/(2\beta J)\langle W_x^2+W_y^2 \rangle$,
  where $W_{x(y)}$ are the winding numbers of the SSE
  worldlines.

  \begin{figure}[h]
\begin{center}
\includegraphics[
bbllx=10pt,bblly=20pt,bburx=340pt,bbury=209pt,%
     width=77mm,angle=0]{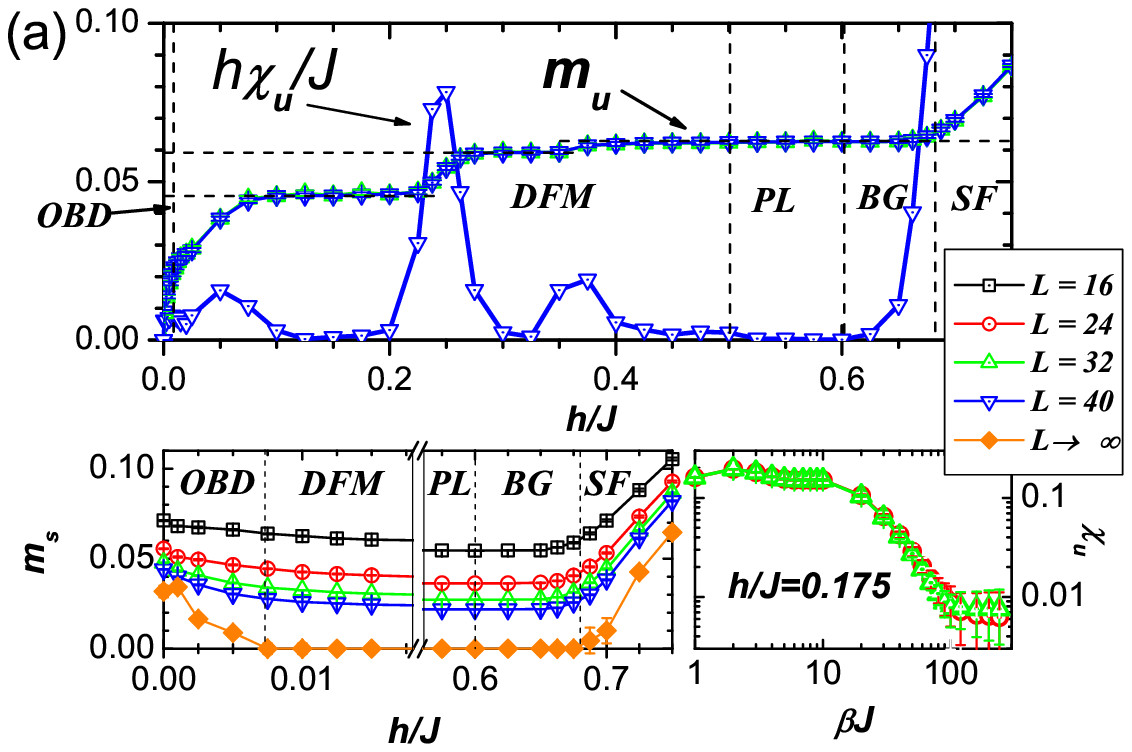}
     \vskip -.2cm
\includegraphics[
bbllx=30pt,bblly=20pt,bburx=300pt,bbury=190pt,%
     width=67mm,angle=0]{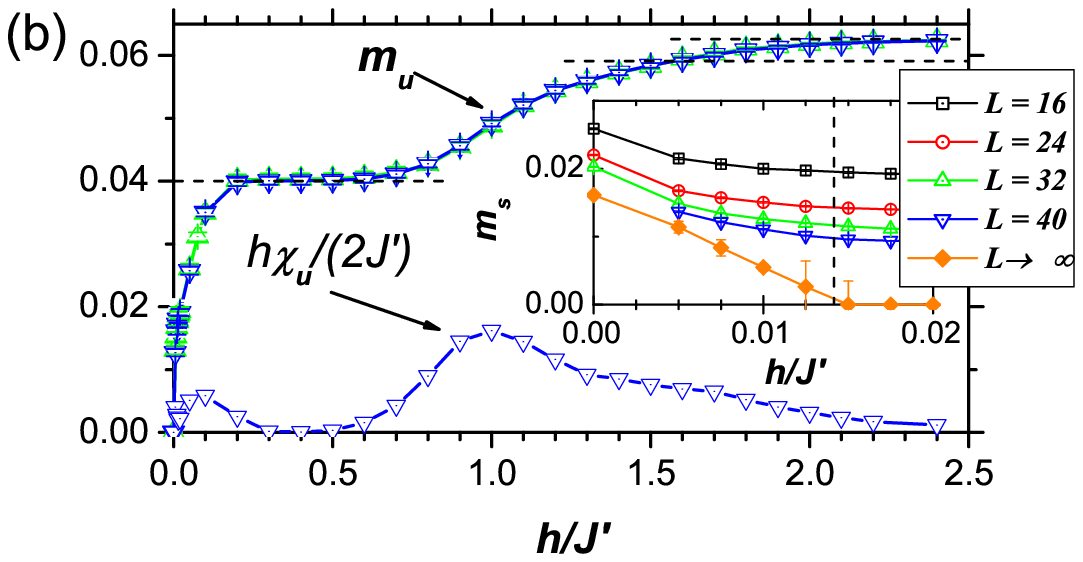}
     \vskip -.2cm
\caption{(Color online) 
Field scan of uniform magnetization, uniform susceptibility, and
staggered magnetization in the diluted system (a) and in the
effective model (b). The lower right panel in (a) shows the
temperature dependence of the uniform susceptibility at $h=0.175J$,
indicating a saturation to a small but
finite value as $T\to0$.
PL=plateau, BG=Bose glass, SF=superfluid.} \label{f.magn}
\end{center}
\null\vspace*{-1.3cm}
\end{figure}

  In the absence of a magnetic field, the effective couplings $J_{ij}$ give
  rise to long-range magnetic order of the network of coupled free moments.
  In particular, for $J'/J$=1/4 and $p$=1/8 we extrapolate
  a staggered magnetization $m_s= 0.032(3)$ 
  in the thermodynamic
  limit. Although the couplings $J_{ij}$ range between 0 and
  $J'$, the average coupling strength responsible for the
  long-range order turns out to be much smaller than
  $J'$ \cite{Roscilde06}. Fig. \ref{f.magn} shows the
  evolution of the ordered moment under application of
  a field. It reveals that the antiferromagnetic order
  is already destroyed at a field $h\ll J'$, namely
  at $h=h_{\rm DFM}= 0.007(1)$. 
  Yet, a striking feature of this disordered phase is
  that the destruction of long-range order is \emph{not}
  accompanied by the full polarization of the FMs,
  as it would ordinarily happen in a homogeneous
  antiferromagnet. At the critical field
  $h_{\rm DFM}$ the uniform magnetization is found
  to be $m_u=0.0208(6)$, 
  much less than the value
  $m_u=pS=1/16$ corresponding to fully polarized FMs, which is
  attained at a much larger field $h_{\rm plateau}/J \approx 0.5$. 
  Consequently the DFM phase,
  appearing between $h_{\rm DFM}$ and
  $h_{\rm plateau}$, is highly unconventional, retaining
  a finite uniform susceptibility and a gapless
  spectrum. These unconventional properties have
  also been observed in the magnetic
  Bose-glass phase of triplet quasiparticles
  living on intact dimers \cite{Fisheretal89,RoscildeH05,Roscilde06,
  Nohadanietal05}. Later we will argue that these two phases
  bear indeed strong analogies, although involving
  different degrees of freedom.

  For $h>h_{\rm plateau}$ the saturation of the
  magnetization of FMs leads to the full restoration
  of a gapped disordered phase due to the field
  \cite{Mikeskaetal04}. Once the field reaches
  the value corresponding to the gap of
  the clean system, $h=\Delta=0.60(1)$,
  a further Bose-glass phase is established
  in which rare clean regions develop
  a local magnetization
  without the appearence of spontaneous order,
  corresponding to localized triplet quasiparticles
  \cite{RoscildeH05, Roscilde06}.
  A delocalization transition of the triplet
  bosons into a superfluid condensate
  corresponds to a further onset of long-range
  transverse order ($m_s>0$) at even higher
  fields \cite{RoscildeH05, Roscilde06}.

    As shown in Fig. \ref{f.magn}, the main features
    of the field dependence
   $m_u$ and $m_s$ for $h<h_{\rm plateau}$ in the doped
   coupled-dimer model of Eq. (\ref{e.ham_dilute})
   are very well reproduced
   by the effective model Eq. (\ref{e.ham_effective}),
   for which we take $\xi_0= 1$ as found by
   simulations at $h=0$ and $p=0$.
   In particular, the fundamental appearence of
   a DFM phase with $m_s=0$ and $\chi_u>0$ is confirmed
   in the effective model. This reveals that
   the FMs are essentially the only degrees of freedom
   responding to a field $h<h_{\rm plateau}$
   in the doped coupled-dimer system.

   \begin{figure}[h]
\null\vskip -1.9cm
\begin{center}
\includegraphics[
bbllx=20pt,bblly=10pt,bburx=370pt,bbury=245pt,%
     width=90mm,angle=0]{
     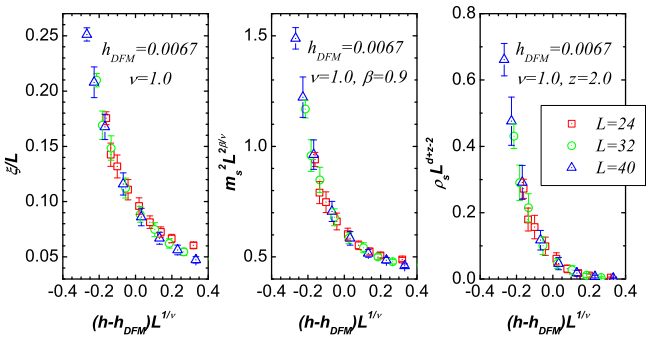}
\caption{(Color online) 
Finite-size scaling of the correlation length, staggered magnetization
and spin stiffness in the vicinity of the quantum critical point
$h_{DFM}$.} \label{f.scaling}
\end{center}
\null\vspace*{-0.8cm}
\end{figure}

   The novel
   quantum phase transition (QPT) between the OBD and the DFM
   phase is studied using finite-size scaling analysis of the
   correlation length, $\xi = L F_{\xi}[L^{1/\nu}\delta h]$,
   of the superfluid density
   $\rho_s = L^{d-2+z} F_{\rho_s}[L^{1/\nu}\delta h]$,
   and of the staggered magnetization
   $m_s = L^{-\beta/\nu} F_{m_s}[L^{1/\nu}\delta h]$,
   where $\delta h = h-h_{\rm DFM}$, as shown in
   Fig. \ref{f.scaling}.
   This allows us to extract the critical exponents
   $z$, $\nu$ and $\beta$. For the original Hamiltonian
   Eq. (\ref{e.ham_dilute}) we find $z=2.0(1)$,
   $\nu=1.0(1)$ and $\beta=0.9(1)$. These estimates
   are also confirmed in the effective FM model.
   The above exponents are fully consistent
   with those of the 2$d$ superfluid-to-Bose-glass
   (SF-BG) QPT previously studied in diluted bilayers
   \cite{Roscilde06}, and
   with the exponents found at the BG-SF transition
   for higher fields for the model Eq. (\ref{e.ham_dilute})
   \cite{Yuetal07}. In particular $z$ is in agreement
   with the general theoretical prediction $z=d$
   \cite{Fisheretal89}, and $\nu$ satisfies the
   fundamental Harris criterion $\nu\geq 2/d$.
   Altogether the present results and those of
   Refs. \onlinecite{Roscilde06,Yuetal07} point towards
   a general SF-BG universality in $d=2$ for order-disorder
   transitions at which the uniform susceptibility
   $\chi_u$ remains finite, corresponding to the
   absence of a gap.

\begin{figure}[h]
\begin{center}
\includegraphics[
bbllx=15pt,bblly=10pt,bburx=310pt,bbury=201pt,%
     width=80mm,angle=0]{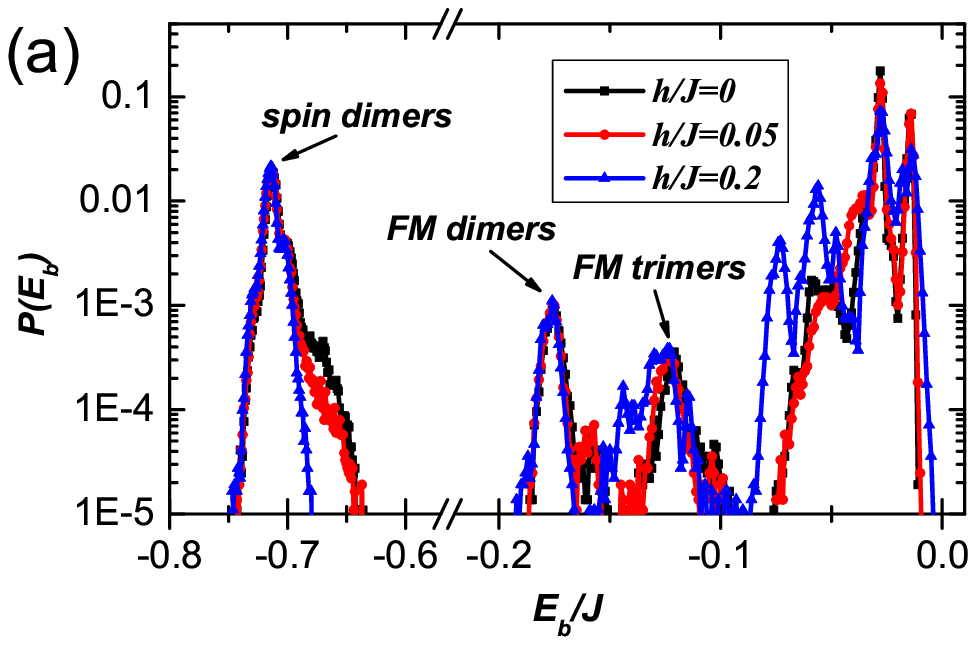}
     \vskip -.3cm
\includegraphics[
bbllx=15pt,bblly=10pt,bburx=310pt,bbury=213pt,%
     width=80mm,angle=0]{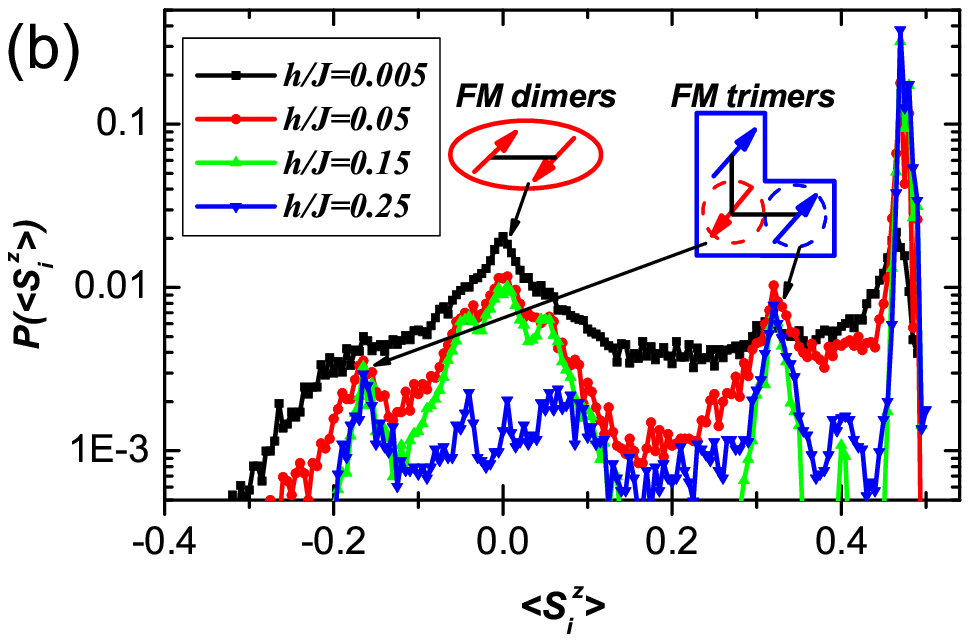}
     \vskip -.5cm
\caption{(Color online) 
Histogram of bond energies (a) and local magnetizations for unpaired
spins (b).} \label{f.histo}
\end{center}
\null\vspace*{-1.2cm}
\end{figure}

   In the DFM phase, the magnetization curves
   of the original Hamiltonian Eq. (\ref{e.ham_dilute})
   and the effective model Eq. (\ref{e.ham_effective})
   show a dramatic feature: beside the large
   pleateau appearing at $h\geq h_{\rm plateau}$, one observes
  the presence of apparent intermediate plateaus at around
  3/4 and $95\%$ 
  of the saturation magnetization.
  A detailed study
  of the temperature-dependent susceptibility
  in this field region reveals that these
  features are actually \emph{pseudo}-plateaus (PPs),
  which retain an extremely small slope (Fig. \ref{f.magn}).
  For both $H$ and $H_{\rm FM}$ the
  first PP extends up to $h\approx 0.7J'$;
  a second PP markedly appears around $h\approx 1.2J'$
  for $H$ (it is rounded off for $H_{\rm FM}$ \cite{Yuetal07});
  the true saturation plateau is only
  attained at $h\approx 2J'$.
  These fundamental features can be
  understood within the picture of strongly
  interacting clusters of FMs in the DFM phase.
  As shown in Fig. \ref{f.DFM}, the zero-field OBD phase
  is essentially inhomogeneous due to the random nature
  of the couplings. A majority of FMs
  are spaced from each other by an average distance
  $\langle r \rangle = p^{-1/d}$, large in the small
  dilution limit, and interact via weak average
  couplings $\langle J_{\rm eff} \rangle \sim p J'$
  ~\cite{Roscilde06}.
  However, fluctuations in the spatial distribution of the impurities
  also lead to small clusters
  of free moments located on neighboring sites,
  and thus interacting with much stronger
  couplings $J'$.
  If antiferromagnetically coupled in even-numbered clusters,
  the strongly interacting
  FMs participate only marginally in the OBD state of the
  system, and have a significant singlet component
  in their ground state wave function. This is
  directly revealed in a histogram of the bond energies (Fig.
  \ref{f.histo}(b)) $E_b = J_b \langle{\bm S}_{1,b} \cdot {\bm S}_{2,b} \rangle$
  where $J_b = J, J'$ and $(1,b)$, $(2,b)$ are the two neighboring
  lattice sites participating in the bond $b$. Beside the peak at
  $E_b\approx -3J/4$, corresponding to singlets on intact dimers,
  a further peak at $E_b\approx -3J'/4$ is observed, corresponding
  to FM dimer singlets, as well as 
  a peak at $E_b\approx -J'/2$ corresponding to FM trimers.

   Applying a magnetic field $h \gtrsim \langle J_{\rm eff} \rangle$
  has clearly the effect of destroying the long-range order of the
  FMs, but at the same time the FM singlets are left intact
  while odd-numbered clusters are not fully polarized
  (Fig. \ref{f.DFM}), with the fundamental
  consequence that the antiferromagnetic order disappears
  but the FMs are far from saturation.
  This is clearly seen in the
  histogram of the local magnetic moments $\langle S_i^z \rangle$
  of the unpaired spins only (Fig. \ref{f.histo}(b)): for small fields,
  a double-peak structure appears with a peak at
  $\langle S_i^z \rangle = S$ corresponding to fully
  polarized FMs, and a strong quantum peak at $\langle S_i^z \rangle = 0$
  corresponding to FM singlets. The large tails for $0 < \langle S_i^z \rangle < S$
  and for $\langle S_i^z \rangle < 0$ come instead from FMs
  in odd-numbered clusters. In fact, we can resolve two more peaks at $\langle S_i^z \rangle =
  1/3$ and $\langle S_i^z \rangle = -1/6$ at larger fields, which
  correspond to partially polarized spins in FM trimers.
  Local FM clusters have widely different
  local gaps to full polarization, both due to
  their geometric structure (dimers,
  trimers, quadrumers, etc.), and to the local field they
  experience from the other FMs. Yet the distribution of rare
  FM clusters clearly assigns dominant statistical weight to the
  dimers, and this simple geometric fact is the
  reason for the appearence of the first PP:
  the magnetization process nearly stops until
  the local gap of the dominant FM dimers is
  overcome at $h \lesssim J'$. Nonetheless,
  the slope remains finite because the FM
  dimers have a distribution of local gaps.
  The magnetization
  value and the field location of the PP can be
  quantitatively related to the statistics of FMs
  clustered in dimers \cite{Yuetal07}. Analogously, one
  can quantitatively associate the second plateau with
  the statistics of the FM trimers \cite{Yuetal07}.
  Higher-order plateaus associated with larger local
  polarization fields should be expected, but they
  cannot be resolved within the given numerical accuracy.

   From the above data a clear picture of the DFM phase
   emerges: in this phase, a majority of the FMs are
   polarized, but antiparallel spins
   exist,
   corresponding to rare FM clusters.
   Upon a spin-to-hardcore-boson
   transformation, these antiparalell spins take the nature
   of bosonic spin-down quasiparticles ($\downarrow$-QPs)
   \emph{localized} on rare regions of the lattice \cite{Yuetal07}.
   This aspect connects with the ordinary picture
   of a Bose glass \cite{Fisheretal89}, and
   it further endows the OBD-to-DFM quantum phase
   transition with the nature of a localization
   transition: in the OBD phase the
   $\downarrow$-QPs form a superfluid
   condensate, which is progressively depleted
   by the applied field (acting as a negative chemical
   potential), up to the point where the $\downarrow$-QPs
   undergo localization into a Bose-glass state,
   losing superfluidity but retaining compressibility,
   which corresponds to a finite uniform susceptibility.

   It is evident from the above results that the
   DFM phase is relevant for experiments
   on site-diluted spin-gapped antiferromagnets
   which display an OBD phase in zero field.
   The fundamental condition for the observation
   of the DFM phase is that the spin gap of the
   pure system be much larger than
   the maximum energy scale of the FM interaction
   (average inter-dimer coupling in weakly coupled
   dimer systems, inter-chain coupling in Haldane
   chains). This condition
   is necessary to ensure that the physics
   of the field response of the FMs is well separated
   in energy from that of the field-induced
   ordered state (for a detailed discussion,
   see Ref. \onlinecite{Roscilde06}). Joint magnetometry
   and neutron scattering measurements at relatively
   low fields should be sufficient to fully pinpoint
   this phase by demonstrating the absence of spontaneous
   order and the finite susceptibility down to zero
   temperature. Furthermore, NMR
   measurements can show the rich structure of the
   distribution of local magnetic moments, similar to
   the histogram of Fig. \ref{f.histo}. The low-field
   location of the DFM phase and its strong physical
   signatures in the magnetic observables
   make it the most
   accessible novel disordered phase in quantum magnets
   with lattice randomness.

   Useful discussions with T. Giamarchi, W. Li, O. Nohadani,
   P. Sengupta, M. Sigrist are gratefully acknowledged.
   It is a pleasure to thank M. Vojta for a remark
   which has sparked the present investigation.
   T.R. is supported by the E.U. through the
   SCALA integrated project. R.Y. and S.H. are supported
   by DOE under grant No. DE-FG02-05ER46240.


\begin{thebibliography}{99}

\bibitem{SandvikS94} A. W. Sandvik \emph{et al.},
Phys. Rev. Lett. {\bf 72}, 2777 (1994).
\bibitem{Matsumotoetal01} M. Matsumoto \emph{et al.},
Phys. Rev. B {\bf 65}, 014407 (2001).
\bibitem{ShastryS81} B. S. Shastry \emph{et al.}, Physica {\bf 108B}, 1069 (1981).
\bibitem{Affleck89} I. Affleck, J. Phys.: Condens. Matter {\bf 1}, 3047 (1989).
\bibitem{Regnaultetal94} L. P. Regnault \emph{et al.}, Phys. Rev. B {\bf 50},
9174 (1994).
\bibitem{DagottoR96} E. Dagotto and T. M. Rice, Science {\bf 235}, 1196 (1987).
\bibitem{Cavadinietal00} N. Cavadini \emph{et al.}, J. Phys.: Condens. Matter {\bf 12},
5463 (2000).
\bibitem{Sasagoetal97} Y. Sasago \emph{et al.}, Phys. Rev. B {\bf 55}, 8357 (1997).
\bibitem{ShenderK91} E. F. Shender \emph{et al.},
Phys. Rev. Lett. {\bf 66}, 2384 (1991).
\bibitem{SigristF96} M. Sigrist \emph{et al.}, J. Phys. Soc. Jpn.
{\bf 65}, 2385 (1996).
\bibitem{Azumaetal97} M. Azuma \emph{et al.}, Phys. Rev. B {\bf 55}, R8658 (1997).
\bibitem{Xuetal00} G. Xu \emph{et al.}, Science {\bf 21}, 419 (2000).
\bibitem{Roscilde06} T. Roscilde, Phys. Rev. B {\bf 74}, 144418 (2006).
\bibitem{Fisheretal89} M. P. A. Fisher \emph{et al.},
Phys. Rev. B {\bf 40}, 546 (1989).
\bibitem{RoscildeH05} T. Roscilde \emph{et al.},
Phys. Rev. Lett. {\bf 95}, 207206 (2005).
\bibitem{Yasudaetal01}  C. Yasuda \emph{et al.},
Phys. Rev. B {\bf 64}, 092405 (2001).
\bibitem{Mikeskaetal04} H.-J. Mikeska \emph{et al.}
Phys. Rev. Lett. {\bf 93}, 217204 (2004).
\bibitem{Laflorencieetal04} N. Laflorencie \emph{et al.},
Phys. Rev. B {\bf 69}, 212412 (2004).
\bibitem{SyljuasenS02} O.F. Sylju\aa sen \emph{et al.},
Phys. Rev. E {\bf 66}, 046701 (2002).
\bibitem{Sandvik02}  A. W. Sandvik, Phys. Rev. B
{\bf 66}, 024418 (2002).
\bibitem{Nohadanietal05} O. Nohadani \emph{et al.},
Phys. Rev. Lett. {\bf 95}, 227201 (2006).
\bibitem{Yuetal07} R. Yu \emph{et al.}, in preparation.

\end{thebibliography}
\end{document}